\documentclass[prl,twocolumn,superscriptaddress,preprintnumbers,a4paper,amsmath,amssymb,showpacs,floatfix]{revtex4}
\usepackage{graphicx}
\usepackage{graphics}
\usepackage{bm}
\usepackage{dcolumn}
\usepackage{color}

\begin{document}


\title{{Low magnetic field phase diagram of UCoGe}}



\author{J.~Vejpravov\'a-Poltierov\'a}\affiliation{Faculty of Mathematics and Physics, Charles University,
DCMP, Ke Karlovu 5, CZ-12116 Praha 2, Czech Republic}

\author{J.~Posp\'i\v{s}il}
\email{jiri.pospisil@centrum.cz}
\affiliation{Faculty of Mathematics and Physics, Charles University,
DCMP, Ke Karlovu 5, CZ-12116 Praha 2, Czech Republic}

\author{J.~Prokle\v{s}ka}
\affiliation{Faculty of Mathematics and Physics, Charles University,
DCMP, Ke Karlovu 5, CZ-12116 Praha 2, Czech Republic}

\author{K.~Proke\v{s}}
\affiliation{Helmholtz-Zentrum Berlin f\"{u}r Materialien und
Energie, Hahn-Meitner Platz 1, D-14109 Berlin, Germany}

\author{A.~Stunault}
\affiliation{Institut Laue Langevin 6 rue Jules Horowitz, BP 156
F-38042 Grenoble Cedex 9 }

\author{V.~Sechovsk\'y}
\affiliation{Faculty of Mathematics and Physics, Charles University,
DCMP, Ke Karlovu 5, CZ-12116 Praha 2, Czech Republic}

\date{\today}

\begin{abstract}
{We propose an alternate scenario for the relation between ferromagnetism (FM) and
superconductivity (SC) in UCoGe at ambient pressure. On the basis of
neutron polarimetry measurement performed in strictly zero field on a
single-crystal at ambient pressure, we claim that the SC phase does exist in spite of
lack of long-range FM order. Moreover, the SC regime dominates to the
paramagnetic (PM) state in small external magnetic fields.
Above the critical SC temperature, $T_{\mathrm{SC}}$ = 0.64~K, the
zero-field state is characterized by a regime with strong FM spin
fluctuations, suggesting the proximity of the FM quantum critical
point. We present an ambient-pressure phase
diagram of UCoGe based on our resistivity, magnetization and
neutron polarimetry results.}
\end{abstract}



\maketitle

A delicate interplay between the quantum criticality and
superconductivity (SC), explicitly in the vicinity of the
ferromagnetic quantum critical point (FM-QCP), represents the
decisive factor in magnetically driven SC
\cite{Levy-Science-2005,Hardy-PRL-2005,Aoki-Nature-2001,Levy-NaturePhysics-2007}.
Recently, Huy et al. reported on UCoGe, which is typified as a
weak ferromagnet that undergoes a subsequent transition at the
critical temperature, $T_{\mathrm{SC}}$ = 0.8~K into a FM-SC
coexistence regime~\cite{Huy-PRL-2007,Huy-PRL-2008}. The magnitude
and anisotropy of the upper critical field suggest $p$-wave
SC and points to an axial SC state with nodes along the
easy magnetization direction (parallel to the $c$-axis),
interpreted in terms of an unusual two-band SC state. UCoGe is
unique among other FM superconductors because of its low
$T_{\mathrm{C}}$ = 2.5~K and a very small spontaneous magnetic
moment of $\sim0.07 ~\mu_B$~\cite{Huy-PRL-2008} directed along the
orthorhombic $c$-axis, indicating the proximity of FM instability.
Although ab-initio calculations~\cite{Divis-PhysicaB-2008}
suggested the existence of an appreciable moment on Co atoms
in zero field, polarized neutron diffraction (PND)
experiments carried out on a single-crystalline sample show that
magnetism is predominantly due to 5f electron states and Co
moments exist only at high magnetic fields~\cite{Prokes-PRB-2010}.
The UCoGe case has been anticipated as the first example of
material with an unconventional SC (USC) stimulated by critical
fluctuations associated with a FM-QCP. In other words, the FM spin
fluctuations (FM-SF), sensitively tuned by an external magnetic
field serve as a
 control parameter mediating the onset of SC.

According to Doniach~\cite{Doniach-proc-1964} and Berk and
Schrieffer~\cite{Berk-PRL-1966}, the phonon-induced $s$-wave SC
in an exchange-enhanced transition metals is suppressed by FM-SF
in the neighborhood of the $T_{\mathrm{C}}$, generally. In contrast,
the Fulde-Furrel-Larkin-Ovchinnikov (FFLO) theory~\cite{FF-PRA-1964,LO-JETP-1965}
allows for a SC state with a spatially modulated
order parameter (OP) coexisting with a long-range FM order in
metallic systems induced by FM impurities.
The spin-correlation theory of the FM state, based on the
interactions mediated by SF between the fermions proposes an
enhancement of the pairing correlations through the FM-SF
\cite{Blagoev-PRL-1999}. The SC phase diagram based on this approach
comprises two SC phases. The $s$-type is established in the FM
state, however the $p$-type state exists in the paramagnetic region
on the border of the FM instability and is expected to vanish at
the QCP.

On the microscopic scale, the critical temperature of the SC
transition depends on the difference of the pairing interaction and
density of states at the Fermi level, respectively, for the
particles with the opposite orientation of the corresponding
pseudospin. In principle, the competition of the stimulating and
suppressing effects on $T_{\mathrm{SC}}$ in FM-SCs determines the phase
diagrams of these unique materials.

The crucial point for the nature of the SC state controlled by
FM-SFs that is relevant for UCoGe is the nature of the zero-field (ZF) state just
above $T_{\mathrm{SC}}$. The question to be answered by studying
high-quality single crystals, is whether the ZF ground
state in UCoGe is indeed a uniform long-range FM or whether UCoGe
represents a system with critical FM-SFs.

In this Letter, we report on the lack of zero-field FM order in
high-quality superconducting single crystal of UCoGe, probed by
bulk measurements and confirmed on the microscopic level using
zero-field neutron polarimetry (ZFNP). Further, we discuss the
anomalous behavior of $T_{\mathrm{SC}}$ in moderate magnetic
fields as indicated by detailed electrical resistivity and
magnetoresistance (MR) measurements. We conclude that the SC in
UCoGe can develop in spite of lack of the long-range uniform FM
order.


Our single crystal of UCoGe was isolated from a large
polycrystalline button as a plate of dimensions: 2 x 1 x 0.5
mm$^{3}$. The crystal was wrapped in Tantalum foil (99.95 \%)
sealed in vacuum 10$^{-7}$~mbar in a quartz tube and annealed at
900 $^{\circ}$C for 10 days. X-ray powder diffraction of small
piece of the crystal found orthorhombic crystal structure (Pnma
space group~\cite{Canepa-JALCOM-1996}) with lattice parameters
\textit{a}~=~684.71~nm \textit{b}~=~420.81~nm
\textit{c}~=~722.63~nm. EDX analysis confirmed chemical
composition of the crystal in the elements ratio 1:1:1. We have
observed spontaneous moment 0.01~$\mu$$_{B}$/f.u.. The
magnetization did not saturate with increasing magnetic field. The
value of the magnetic moment at temperature 1.8 K and magnetic
field 7~T was 0.17~$\mu$$_{B}$/f.u.. These results together with
indirect parameter \textit{RRR}~=~20 assured us of the high
quality of the studied crystal.

The bulk measurements were performed in a commercial SQUID
magnetometer and a Physical Property Measurement System (PPMS)
device. The electrical resistivity experiments were carried out
down to 0.35 K with experiment arrangement
\textit{I}$\parallel$\textit{B}$\parallel$\textit{c}. Before each
ZF and zero-field-cooled (ZFC) experiment we applied a standard
demagnetization procedure in order to obtain the lowest remanent
field possible. Control measurement using pure indium showed that
the remanent field was of the order of 0.1 mT. We repeated all the
low-field experiments several times.

In order to probe the ZF ground state of UCoGe on a microscopic
level we have performed a ZFNP experiment, a special type of PND.
In the simplest PND, one records Bragg reflection intensities
$I^{\pm}(\bf{Q}) \propto$ $|F_N$($\textbf{Q}$) $\pm$ $F
_M$($\textbf{Q}$)$|^2$, where the $+$ and $-$ sign refer to up and
down polarization directions of the incoming neutron beam and
calculates the so-called flipping ratios (FR)
$FR(\textbf{Q}$)=$I^{+}({\bf Q})/I^{-}(\bf{Q})$. The crystal
$F_N$($\textbf{Q}$) and magnetic $F _M$($\textbf{Q}$) structure
factors depend on the nature, state and spatial distribution of
the constituent atoms and magnetic moments (and their directions),
respectively. It is evident that this technique makes use of the
interference term between previously determined
$F_N$($\textbf{Q}$) and $F_M$($\textbf{Q}$). For Bragg reflections
of either pure nuclear or magnetic origin absence of the
interference term leads to $FR$ = 1.00. From mathematical
formula~\cite{Lelievre-Berna-PhysicaB-2007-1} describing the
influence of individual nuclear and magnetic contributions to
measured FRs, it follows that off-diagonal elements, if different
from zero, indicate possibly a complicated non-collinear magnetic
order. In the case that UCoGe exhibits FM order, it is expected
that the sample would consists from magnetic domains forcing these
elements to be close to zero. On the contrary, diagonal elements
would be in this case lower than 1.00 due to beam depolarization.

The experiment was carried out using CRYOPAD
polarimeter~\cite{Lelievre-Berna-PhysicaB-2007-1} in D3
diffractometer at the ILL with polarized neutrons, $\lambda$ =
0.825 \AA. All the measured data have been consequently corrected
for both polarizer and analyzer efficiencies.

We have determined the FR matrix for several
reflections at different temperatures well above and below the
suggested Curie temperature. The philosophy of the experiment was
simple and straightforward - if there is any long-range FM order
present in the sample, significant deviations from 1.00 in the
diagonal components of the FR matrix below the $T_{\mathrm{C}}$ must
be observed. To optimize the resulting statistical error for fixed
measurement time we used the approach described in detail in ref.
\cite{Lelievre-Berna-PhysicaB-2007-2}.


\begin{figure}
\includegraphics*[scale=0.40,angle=0]{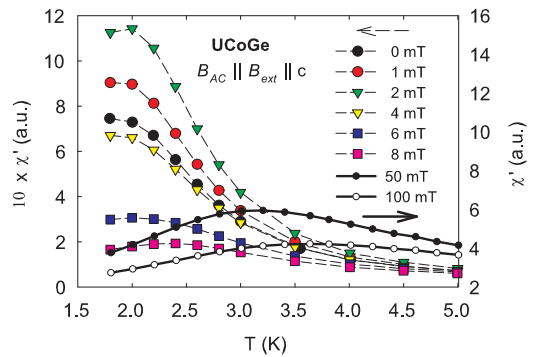}
 \caption{(Color online) Low-temperature a.c. susceptibility (real
part, $\chi$') for external d.c. fields, $B_{\mathrm{ext}}$ applied
parallel to the $c$-axis and to the a.c. driving field
($B_{\mathrm{ac}}$ = 0.1 mT).} \label{Fig1}
 \end{figure}

In the original paper by Huy et al.~\cite{Huy-PRL-2007}, the
formation of the FM state was manifested by the appearance of a
maximum at around 2.5 K on the temperature dependence of the AC
susceptibility (see Fig.~\ref{Fig1}). Our single crystal
experiments under similar conditions (the AC magnetic field,
$B_{\mathrm{ac}}$ = 0.1~mT, and the external DC field applied
along the easy axis) revealed no clear anomaly around
$T_{\mathrm{C}}$ in the temperature dependence of the real part of
the AC susceptibility, $\chi$' when measured in ZF. However, an
abrupt enhancement of the signal with a symmetric maximum with
increasing d.c. magnetic field up to 2~mT is observed. With
further increasing field, the signal becomes continuously
depressed, the maximum becomes broader and continuously extends to
higher temperatures. The AC susceptibility results suggest, that
in ZF, at temperature $\textit{T}$ $\sim$ 2.0~K, a paramagnetic
state with strong critical anisotropic FM fluctuations is
established in our sample.

\begin{figure}
\includegraphics*[scale=0.40,angle=0]{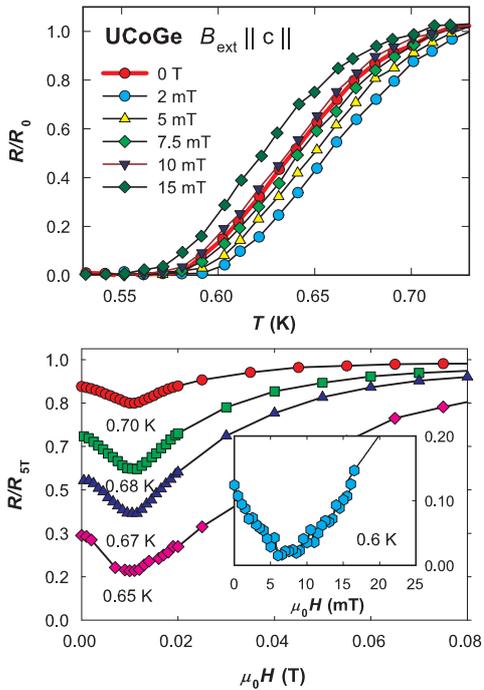}
\caption{(Color online) (a) Low-temperature electrical resistivity
normalized to the value at 0.75~K measured with electrical current
and magnetic fields applied along the $c$-axis. (b) Low
temperature magnetoresistance measured along the $c$-axis
normalized to values obtained at field of 5~T applied along the $c$-axis. In the inset, measurement at 0.6~K is shown. } \label{Fig2}
\end{figure}

The coexistence of the FM-SF and the USC state was subsequently
explored by detailed measurements of the electrical resistivity
and MR as shown in Fig.~\ref{Fig2}a. The bulk SC state has been
confirmed by measuring of the Meissner effect, as well. The
principal observation is a transient increase of the
$T_{\mathrm{SC}}$ under applied magnetic field in low fields.
First, the $T_{\mathrm{SC}}$ (determined as the inflection  point
on the resistivity curve) moves from 0.64~K to 0.66~K as the field
is increased from $B$ = 0~T to $B$ = 2~mT, respectively. It moves
down, to the initial zero-field value, for a field of 10~mT, and
finally monotonously decreases with increasing magnetic field
applied, as expected for any SC system. This unusual field
dependence is supported by the MR measurements at temperatures in
the vicinity of the $T_{\mathrm{SC}}$. The MR curve exhibits a
clear minimum in a field of 10~mT, which becomes continuously
smeared out with increasing temperature. Although all these data
support the idea that in the absence of a magnetic field UCoGe
does not posses static long range FM order; it does not prove
unambiguously that the FM is really missing. Other order types
that include even ferrimagnetism as suggested by ab-initio
calculations~\cite{Divis-PhysicaB-2008} cannot be principally
excluded. The crucial proof, however, is provided by the above
described ZFNP experiment.

In Table~\ref{tab:table1} we list as an example FR matrices
determined for the (202) and (301) Bragg reflections at two
different temperatures, one well above and the other well below
the suggested ferromagnetic phase transition $T_{\mathrm{C}}\sim$
2.5 K. First we note that all off-diagonal elements are negligible
and within error bars equal to zero. Second, all diagonal elements
are close to 1.00. This is not surprising for data taken at 15 K
as at this temperature is UCoGe not magnetically ordered. However,
diagonal elements are equal to 1.00 below the proposed
$T_{\mathrm{C}}$ and even below $T_{\mathrm{SC}}$ (not shown). The
same observation has been made also for all other measured Bragg
reflections suggesting that the ZF conditions that are necessary
prerequisite for conservation of the neutron polarization within
the CRYOPAD zero field chamber are fulfilled in our case at all
temperatures. This is possible only if UCoGe is not ferromagnetic
in the whole temperature range. This result is in clear contrast
to ($\mu$-SR)~\cite{DeVisser-PRL-2009} experimental results by de
Visser et al. for a polycrystalline sample. To explain the
apparent controversy we recall that polycrystalline UCoGe samples
have often larger magnetization than single crystals that need to
be carefully heat-treated. Moreover, positive muons sense local
moments on another time window than neutrons and other local
probes. The $^{59}$Co NQR [18] measurements also on
polycrystalline sample led to conclusion that UCoGe is in
self-induced vortex state having about 30\% of the volume
superconducting. Let us also note that the internal field
($B_{\mathrm{loc}}$) deduced from $\mu$-SR [17] is 15 mT. This
field is about four orders of magnitude larger than the residual
field inside the CRYOPAD chamber. So even if the sample would
consist from FM domains, resulting stray field would destroy the
neutron polarization. When considering the values of the
relaxation rate ($\Delta$) $\sim$ 0.3 ms$^{-1}$ and internal field
$B_{\mathrm{loc}}$ $\sim$ 15 mT in UCoGe [17], the specific
$\mu$-SR time window ranges from 10$^{-10}$ to 10$^{-7}$ s, which
fairly covers that of the PN scattering. Therefore, the sample
aggregation state and/or interference of the probe with the
sample, respectively, play an important role in interplay of
FM-SC.

\begin{table*}
\caption{\label{tab:table1}The list of flipping ratios for (202) and
(301) Bragg reflections measured on UCoGe single crystal above and
below the suggested ferromagnetic phase transition
$T_{\mathrm{C}}\sim$ 2.5~K. In parenthesis the statistical errors
according to~\cite{Lelievre-Berna-PhysicaB-2007-2} are shown.}

\begin{ruledtabular}
\begin{tabular}{c|c|c|ccc|c|ccc}
T~~& $hkl$~&$P_{\mathrm{in}}$ & & $P_{\mathrm{out}}$ & &$hkl$~& &$P_{\mathrm{out}}$ & \\
         ~(K)~~~&~~~~&   & $+(x)~$ &  $+(y)~$ & $+(z)~$ & & $+(x)$ &  $+(y)$ & $+(z)$ \\
 \hline
      15.0~~&    &$+(x) $~~~&  1.018(9) &-0.002(11)&-0.005(11)&    & 0.97(2)&-0.05(2)&-0.04(2)  \\
      15.0~~&202~~~&$+(y) $~~~& -0.030(11)& 1.007(9) & 0.024(11)&301~~~& 0.00(2)& 1.02(2)&    not measured     \\
      15.0~~&    &$+(z) $~~~&  0.005(11)& 0.013(12)& 0.999(11)&      & 0.01(2)&    not measured   & 1.00(2)  \\
           \hline
      1.15~~&    &$+(x)$~~~&  1.009(11)& 0.033(13)& 0.002(13)&     & 1.00(2)&-0.01(2)&-0.01(3)  \\
      1.15~~&202~~~&$+(y)$~~~& -0.007(14)& 1.019(12)& 0.011(15)&301~~~&-0.02(3)& 0.98(2)& 0.01(3)  \\
      1.15~~&    &$+(z)$~~~&  0.012(15)& 0.01(2)  & 1.002(13)&      &-0.03(3)& 0.02(3)& 1.05(3)  \\
       \end{tabular}
\end{ruledtabular}
\end{table*}

\begin{figure}
\scalebox{0.3}{\rotatebox{270}{\includegraphics{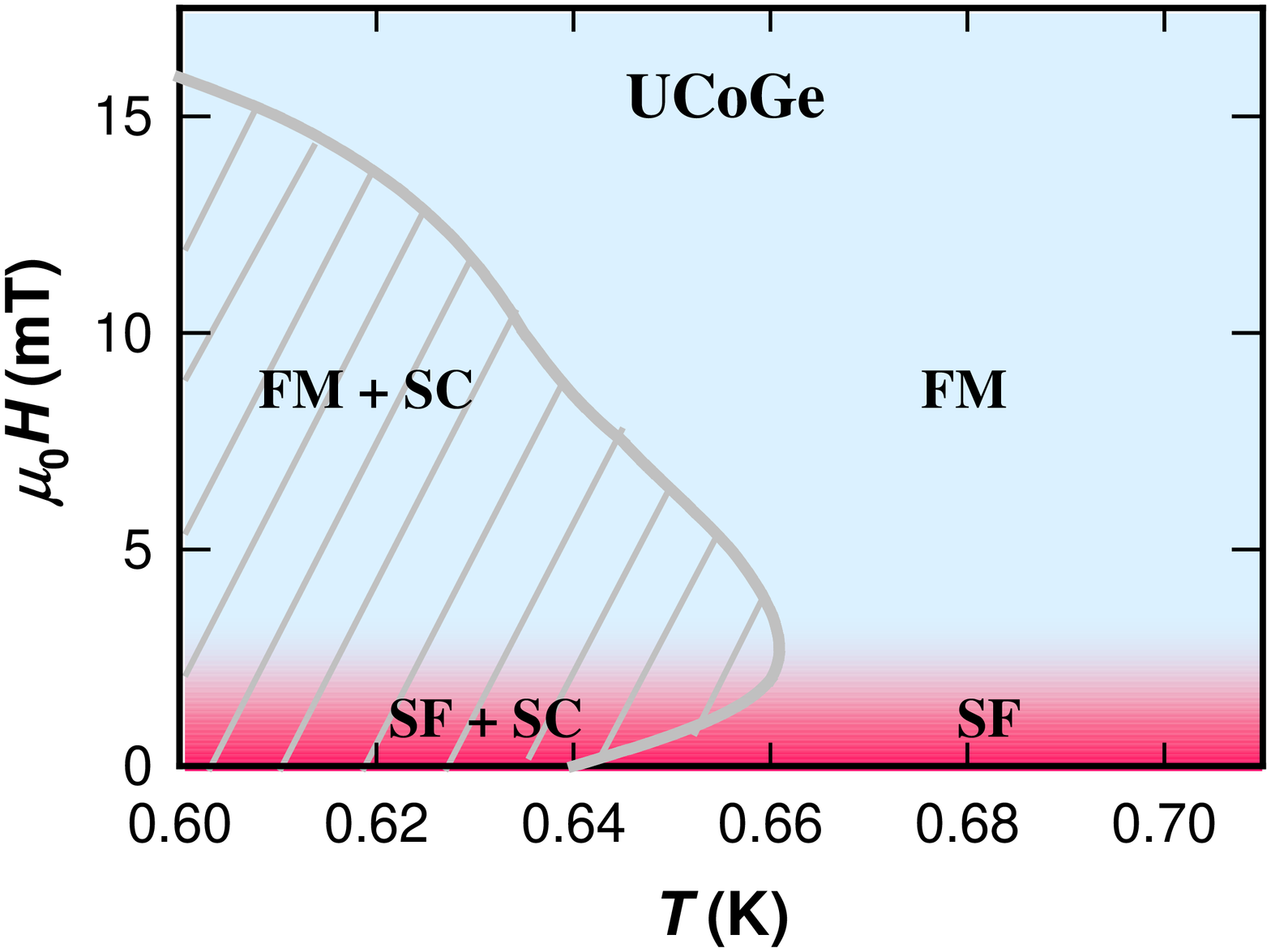}}}
 \caption{ Suggested magnetic phase phase diagram of UCoGe single crystal for magnetic field applied along the $c$-axis.}
 \label{Fig3}
 \end{figure}

On the basis of our experimental results, we propose an
approximate low-temperature magnetic phase diagram in
Fig.~\ref{Fig3}. There are two essential questions about the proposed
phase diagram. The first is a competition of FM-SF against the robust FM
state. The second point is the classification of the two generally
different SC states related to the phase SF-SC and FM-SC,
respectively. The major controversy points to the
nature of the low-field SC state related to the phase SC-SF. Let us
propose two fundamentally diverse scenarios. In the first, the
two SC states are of different symmetries. In zero-field, the order
parameter is expected to be isotropic resulting in the only allowed
s-type SC, but when we reach the robust FM state, the
anisotropy of the Fermi surface is fully developed and the
unconventional p-type can be formed. The proposed mechanism can be
supported by the theory of Blagoev~\cite{Blagoev-PRL-1999}, which claims, that
for weakly FM metals the coexistence of longitudinal SF and a
gapless Goldstone mode allows formation of both the s-wave and the p-wave \cite{Roussev-PRB-2001} type SC. The crucial
discrepancy is, that the p-type should occur in the paramagnetic
limit and the s-type in the FM. On the other hand, we have observed
significant changes on the FM-related a.c. susceptibility, which can
be explained by a change in the anisotropy and dynamics of the SF.
Isotropic SF in the vicinity of the cross-over point (potential QCP)
therefore suggest possibility of the two SC states with a different
symmetry. We can also consider the effect of FM domains in USC, so
the SC phases have the same symmetry of the OP. For an orthorhombic
system, there are four classes, and two pairs of equivalent
co-representations, which in general have a different
$T_{\mathrm{SC}}$. The equivalent co-representations correspond to
the same SC phase, but each of them describes the sub-state in
magnetic domains \cite{Mineev-PRB-2002} with the magnetic moment with orientation parallel
or perpendicular to the magnetic field direction. We may observe
either a crossover between the two states of the two
co-representations, or homogenize the FM state by the applied
magnetic field, so the proper p-type SC state \cite{Scharnberg-PRL-1985} is established within
one of the co-representations. In general, there is no reason for
competition of the two in general analogues SC phases. Therefore we
propose, that the anomalous behavior of $T_{\mathrm{SC}}$ in the SF-SC phase
 is not due to the crossover between two different
co-representations of the same group, but rather due to the magnetic field
influences on the critical FM-SF, which is the governing factor
conditioning the phase line of the SC state.

In conclusion, we have investigated relation of SC
and FM in UCoGe in strictly zero-field conditions. We
have observed: 1. fully-robust SC state in spite of no long-range FM
order in zero magnetic field and ambient pressure, and 2.
enhancement of the $T_{\mathrm{SC}}$ under applied small
magnetic fields in the vicinity of the SC and FM phase lines. We propose a B-T
phase diagram comprising the critical phenomena in low magnetic
fields on the border of the SC region. Finally, our results
suggest a crossover between two SC phases, in analogy to the
observations at elevated pressures by Slooten et al.~\cite{Slooten-PRL-2009}.

We acknowledge ILL for allocation of beamtime for performing the
neutron diffraction experiment and the ILL staff for support
during the experiment. This work is a part of the research plan
MSM 0021620834 that is financed by the Ministry of Education of
the Czech Republic. It was also supported by the grant no.
202/09/1027 of the Czech Science Foundation and by the grant no.
154610 of the Grant Agency of Charles University. JPV acknowledges
postdoctoral project no. 202/08/P006 (G.A. of the CR)


\end{document}